\documentclass[12pt,preprint]{aastex}
\begin{document}

\title{ANALYSIS OF 3D MHD INSTABILITY OF ANTISOLAR LATITUDINAL
DIFFERENTIAL ROTATION IN F, G AND K STARS}
\author{Mausumi Dikpati$^1$ and Paul S. Cally$^{1,2}$}
\affil{1. High Altitude Observatory, National Center for Atmospheric
Research \footnote{The National Center
for Atmospheric Research is sponsored by the
National Science Foundation. },
3080 Center Green, Boulder, Colorado 80301; dikpati@hao.ucar.edu}
\affil{2. Monash Centre for Astrophysics, School of
Mathematical Sciences, Monash University, Clayton, Victoria,
AUSTRALIA 3800; paul.cally@monash.edu}

\begin{abstract}
Motivated by observations that only a very few stars
have been found to have antisolar differential rotation,
much weaker in amplitude than that of the Sun, we analyze
the stability of antisolar and solar type latitudinal
differential rotations in the tachoclines of typical
F,G and K stars. We employ two three-dimensional thin
shell models, one for a Boussinesq but non-hydrostatic system,
the other for a hydrostatic but non-Boussinesq system.
We find that, in general, the combination of toroidal
field band and differential rotation is more unstable,
and unstable for lower toroidal fields, for antisolar
than for solar-type differential rotation. In the
antisolar case, the instability is always found
to weaken the differential rotation, even if the primary
energy source for the instability is the magnetic field.
This favors surface antisolar differential rotations in stars
being weaker than solar types, if the instability in the
tachocline is felt at the surface of the star. This is
most likely to happen in F stars, whose convection zones
are much thinner than they are in G and K stars. This effect
could help explain why the antisolar differential rotations
that have been found are very weak compared to the Sun. 
\end{abstract}

\section{Introduction}

The internal rotation pattern of a star has most extensively been
studied observationally and theoretically for our nearest G-star,
the Sun. For the surface of the Sun, \citet{carrington1858} showed that
the equator rotates faster than the pole, by systematically monitoring
the rotation of sunspots over the 11-year cycle. In the 20th century,
the Sun's positive pole-to-equator differential rotation was established
from Doppler shift measurements \citep{delury1939, plaskett1959}. A
review of early observations of solar surface differential
rotation can be found in \citet{gilman1974}. Quantitative analyses of
long term Mount Wilson Observatory data indicates that the equator
rotates about 132 nHz faster than the pole \citet{uetal1988} at the
surface. Helioseismic measurements revealed that the Sun's
positive pole-to-equator differential rotation pattern persists
from the surface down to the core-envelope interface, through the
bulk of the convective envelope -- the details can be found in
an extensive review by \citet{tcmt2003}.

Non-axisymmetric instability of solar-like, positive pole-to-equator
differential rotation in the presence or absence of magnetic fields
have been extensively studied in global 2D, quasi-3D and 3D thin-shell
models \citep{watson1981, dk1987, gf1997, dg1999, cdg1999, garaud2000,
dg2001, cally2001, gd2002, cdg2003, dgr2003, cally2003, dcg2004,
asr2005, asr2007, mgd2007, gdm2007}. Most of these calculations were 
applied to the solar tachocline, the interface between the solar 
convection zone and interior, where the stratification is thought to 
be either slightly subadiabatic, as in the overshoot layer where
convection penetrates from above, or much more subadiabatic, as in the 
radiative layer immediately below the overshoot layer. But differential 
rotation anywhere in a star may be subject to the instabilities described 
in the above references.

Axisymmetric instability of latitudinal differential rotation can not 
occur in 2D HD or MHD (see discussions in \citet{gf1997, dg1999, cdg2003}) 
and is hard to excite in quasi-3D shallow-water models for which about 
2 million Gauss peak toroidal magnetic field is required. In a 3D thin-shell 
model, however, axisymmetric instabilities can be excited with a much lower 
toroidal field of about 5,000 Gauss. A few recent calculations have been 
done to investigate axisymmetric instabilities in solar tachocline 
\citep{cdg2008, dgcm2009, hc2009}. These studies show that latitudinal 
differential rotation of up to 18\% pole-to-equator amplitude will be 
hydrodynamically stable in 2D, but is unstable in 3D or in the presence 
of magnetic fields.

Various observations, including light curves of rotating starspots
\citep{mg2003}, spectroscopic measurements \citep{rs2003}, long term
changes in Ca II H and K fluxes \citep{dsb1996}, and Doppler imaging
\citep{bcjd2000} indicate the existence of differential rotation in many
stars. Among the stellar population of F, G and K stars, solar-like
positive pole-to-equator differential rotation, as well as antisolar
pattern in which the equator rotates slower than the pole, are observed.
For example, multiwavelength studies of G0-G5V stars by \citet{mg2003},
and Zeeman-Doppler imaging by \citet{jd2009} have revealed the existence
of solar-like (positive) and antisolar (negative) pole-to-equator
differential rotation.

For the spot-dominated late type stars, inversions of light curves for
images of dark starspots on the surface reveal the differential
rotation pattern of K-stars; so far all of them were found to have
solar-like profiles \citep{rhvh2011}. Antisolar differential rotation
has been found among F and G stars, but occurs much less frequently than
solar-like differential rotation. Furthermore, the amplitude of
antisolar differential rotation in those stars are also found to be
smaller than solar type pole-to-equator differential rotation.

Theoretical modeling has enabled us to understand how the solar-like
positive pole-to-equator differential rotation is formed in the solar
and in stellar convection zones \citep{gm1986,rempel2005}. Stars with 
deeper convection 
zones are likely to have broader differential rotations with latitude 
\citep{gilman1979}. Except for anomalously weak rotators, such stars 
are likely to have equatorial acceleration like the Sun does. On the 
other hand, stars with shallow convection zones, like F stars, may not 
have broad differential rotation, but rather a much more structured 
pattern. Even if they rotate much faster than the Sun, if the turnover 
time in their shallow convection zone is short enough, they could have 
antisolar differential rotations.

Hydrodynamic and magnetohydrodynamic instability of solar-like 
differential rotation in stars has also been explored 
\citep{ks1982,uss1996, spruit1999, spruit2002, braithwaite2006}. But to 
our knowledge for the case of antisolar differential rotation neither 
detailed theory of formation nor the stability analyses have been 
carried out yet.

The aim of this paper is to analyze the instability of antisolar
differential rotation in order to answer the following questions.
(i) Why have only about a dozen antisolar stars been found?
(ii) Is antisolar differential rotation more unstable than the
solar-like pattern, and could that be the reason? (iii) What is the
limiting amplitude of a negative pole-to-equator differential rotation
that can be stable? (iv) Why has no antisolar K-star been found yet?

The observations of stellar rotation are made at the surface of
a star, and we do not know the differential rotation pattern
in their tachoclines. Using the analogy that the Sun's
positive pole-to-equator surface latitudinal differential rotation
persists down to the tachocline, we assume that the stellar tachoclines
reflect the same sign of the pole-to-equator latitudinal
differential rotation as observed at their surfaces. The differential
rotation formed in the stellar convection zone imposes an upper boundary
condition on the tachocline and provides one source for the energy
needed to drive instabilities there.

But the precise relation between stellar rotation at the surface
and in the stellar tachocline may be more complex, and depend
on such factors as the thickness of the stellar convection zone. In the 
case of the Sun, and probably K stars, the inertia of the convection zone 
is so large compared to the tachocline below it that it is unlikely that 
instability in the tachoclines of such stars could significantly change 
the surface differential rotation pattern. \citet{rempel2005} did show 
that thermal effects in the solar tachocline could modify significantly 
the rotation contours of the convection zone, moving them away from being 
constant on cylinders as global convection theory tends to create. But 
with the thin, low density convection zones of F stars, instability in 
the tachocline could have a much stronger feedback on the surface 
rotation, resulting in a lower overall differential rotation for F-stars, 
including at their photospheres. This is a reason to study tachocline 
instabilities in stars as one determinant of their surface differential 
rotations.

We will use for our calculations a 3D thin-shell model of the
solar/stellar tachocline to analyze axisymmetric and nonaxisymmetric MHD
instability of the pole-to-equator latitudinal differential rotation for
a wide range of positive and negative amplitudes. We do the MHD instability
problem as opposed to the purely HD problem because it is likely that most 
stellar tachoclines contain substantial toroidal fields. Also, in the
previous calculations on magneto-shear instabilities by us and many
others, as referred earlier in this section, major findings in 2D, quasi-3D
and 3D models were that the magneto-shear instability exists for a wide 
range of toroidal field bands of latitudinal widths of above $2.7^{\circ}$, 
located at a wide range of latitudes above $\sim 13^{\circ}$, and for a 
wide range of peak field strength from a few hundred Gauss and above. 

Magneto-shear instabilities occur for differential rotation amplitudes 
that are characteristic of both radiative and overshoot tachoclines. 
In the case of the Sun, surface observations can give some guidance 
about the bandwidth of the tachocline toroidal field bands to be of 
$6 - 10^{\circ}$, if active regions are being produced from them 
(see \citet{zwaan1978}). Rising flux tube studies (see, for example, 
\citet{ffd1993}) indicate that these bands also perhaps spend their 
maximum time in the latitude range of $0^{\circ} - 45^{\circ}$. So, 
in order to pick a typical case for the Sun and solar-like stars for 
investigating magneto-shear instabilities with solar and antisolar 
differential rotation, we consider in this study a $10^{\circ}$ 
toroidal band located at $30^{\circ}$ latitude, and explore the 
details in the parameter space of solar/antisolar differential 
rotation amplitude and toroidal field strength. 

\section{Physical Context and Parameter Choices}

Full 3D and 3D thin shell models of the solar tachocline have been
developed by several authors \citep{zls2003, cally2003, asr2007,
gdm2007, mgd2007} to explore the global instability of tachocline
latitudinal differential rotation in the presence of toroidal
magnetic fields. For investigating global instability of antisolar
differential rotations in F, G and K stars, we perform the analyses
in a 3D thin-shell model of the solar/stellar tachocline, but using
two different approaches, namely (i) a hydrostatic, non-Boussinesq
system and (ii) a nonhydrostatic, Boussinesq system. For (i), we start
from the 3D thin shell perturbation equations (16)-(25) of
\citet{gdm2007}, linearized about a reference state containing
differential rotation ($u_0$), toroidal field ($a_0$), pressure ($p_0$)
and temperature ($\theta_0$). In latitude, longitude, depth coordinate
system ($\phi$, $\lambda$, $z$) the latitudinal and vertical force
balance of the reference state can be described in the following equations:
$$({\alpha_0}^2 -{\omega_0}^2)\,\cos\phi\,\sin\phi = {\partial \pi_0
\over \partial\phi}, \quad\eqno(1)$$
$${\partial \pi_0 \over \partial z} = -G^{1/2} {\delta_S \over \delta_E}
\rho_0
=-{\delta_S \over \gamma} p_0 + G^{1/2}
\theta_0, \quad\eqno(2)$$
$$p_0 = \gamma G^{1/2} \left[{\delta_E}^{-1} \rho_0 + {\delta_S}^{-1}
\theta_0 \right], \quad\eqno(3)$$
$$\pi_0 = p_0 + {{\alpha_0}^2 \,\cos^2\phi \over 2}. \quad\eqno(4)$$
These are equations (11-14) in \citet{gdm2007}. We specify $\alpha_0$
and $\omega_0$, which are respectively the toroidal field and differential
rotation in angular measure. Assuming that the perturbation variables are
of the form,
$u,\,v,\,w,\,a,\,b,\,c,\,\rho,\,p,\,\pi,\,\theta\,\sim e^{im(\lambda
-\sigma t)},$,
in which $m$ is the longitudinal wave number, and representing each variable
$u$ as $u=u_c(\mu) \cos(n{\cal \pi} z) + u_s(\mu) \sin(n\pi z), {\rm for}
0\le z\le1,$ in order to separate $z$ from $\mu$, we obtain the
following linearized first order perturbation equations which we will
solve using parameters chosen to represent F, G and K stars.

The five equations from coefficients of $\cos(n\pi z)$ are:

$$ im(\omega_0 - \sigma)u_c -im\alpha_0\,a_c -2\mu (\omega_0 v_c -
\alpha_0 b_c) \hspace{3cm}
$$
$$ \hspace{2cm} + {(1-\mu^2)}\left({\partial\omega_0 \over \partial\mu}v_c
-{\partial\alpha_0 \over \partial\mu}b_c \right) + {im \over
{(1-\mu^2)}^{1/2}}\pi_c=0, \quad\eqno(5)$$

$$i m(\omega_0 - \sigma)v_c -im\alpha_0 b_c -2\mu (\alpha_0 a_c -
\omega_0 u_c) + {(1-\mu^2)}^{1/2} {\partial \pi_c \over \partial \mu}=0,
\quad\eqno(6)$$

$$ i m(\omega_0-\sigma)a_c - i m\alpha_0 u_c + v_c (1-\mu^2) {\partial
\alpha_0 \over \partial\mu} -b_c (1-\mu^2) {\partial\omega_0 \over
\partial\mu} =0, \quad\eqno(7)$$

$$i m(\omega_0 - \sigma)b_c- i m\alpha_0 v_c=0, \quad\eqno(8)$$

$$i m(\omega_0 - \sigma) \left[-G^{-1/2}
{(n \pi)}^2 \pi_c +{\delta_S \over \gamma {G}^{1/2}}(n\pi) \pi_s
- {\delta_S \over \gamma G^{1/2}} {(1-\mu^2)}^{1/2} \alpha_0 (n\pi) a_s
\right] \hspace{2cm}
$$
$$\hspace{1cm}
+ {(1-\mu^2)}^{1/2}{\partial \theta_0 \over \partial \mu}(n\pi) v_s
- G^{1/2} \left[{im \over {(1-\mu^2)}^{1/2}} u_c + {\partial
\over \partial\mu}\left({(1-\mu^2)}^{1/2} v_c \right)\right]
=0, \quad\eqno(9)$$

\noindent and the five from coefficients of $\sin(n\pi z)$ are:

$$ im(\omega_0 - \sigma)u_s -im\alpha_0\,a_s -2\mu (\omega_0 v_s -
\alpha_0 b_s) \hspace{3cm} $$
$$ \hspace{2cm}
+ {(1-\mu^2)}\left({\partial\omega_0 \over \partial\mu}v_s
-{\partial\alpha_0 \over \partial\mu}b_s \right) + {im \over
{(1-\mu^2)}^{1/2}}\pi_s=0, \quad\eqno(10)$$

$$i m(\omega_0 - \sigma)v_s -im\alpha_0 b_s -2\mu (\alpha_0 a_s -
\omega_0 u_s) + {(1-\mu^2)}^{1/2} {\partial \pi_s \over \partial \mu}=0,
\quad\eqno(11)$$

$$ i m(\omega_0-\sigma)a_s - i m\alpha_0 u_s + v_s (1-\mu^2) {\partial
\alpha_0 \over \partial\mu} -b_s (1-\mu^2) {\partial\omega_0 \over
\partial\mu} =0, \quad\eqno(12)$$

$$i m(\omega_0 - \sigma)b_s- i m\alpha_0 v_s=0, \quad\eqno(13)$$

$$i m(\omega_0 - \sigma) \left[G^{-1/2}
{(n \pi)}^2 \pi_s +{\delta_S \over \gamma {G}^{1/2}}(n\pi) \pi_c
- {\delta_S \over \gamma G^{1/2}} {(1-\mu^2)}^{1/2} \alpha_0 (n\pi) a_c
\right] \hspace{2cm}
$$
$$\hspace{1cm}
+ {(1-\mu^2)}^{1/2}{\partial \theta_0 \over \partial \mu}(n\pi) v_c
+ G^{1/2} \left[{im \over {(1-\mu^2)}^{1/2}} u_s + {\partial
\over \partial\mu}\left({(1-\mu^2)}^{1/2} v_s \right)\right]
=0. \quad\eqno(14)$$

Equations (5-14) are the equations (33-42) of \citet{gdm2007} and the
detailed derivation is given there. The nonhydrostatic-Boussinesq 
system behaves very similarly to the hydrostatic-nonBoussinesq system
except for $G \approx 0$. For Boussinesq system, we use the 
formulation and the solution method of \citet{cally2003}. 

Within these dimensionless equations two parameters, $G$ and $\delta_s$,
appear frequently and must be evaluated for typical F, G and K stars
we are considering. $\delta_s = H/H_p$, in which $H$ is the
thickness of the stellar tachocline and $H_p$ is the pressure
scale height at tachocline depth. This parameter is a measure
of the importance of the density decline with radius within the
tachocline.

The parameter $G= g |\nabla -\nabla_{ad}| H^2/2(r_0 \omega_c)^2 H_p$, in
which $g$ is gravity at tachocline depth, $|\nabla-\nabla_{ad}|$ is the
fractional departure of the temperature gradient from the adiabatic
value, $r_0$ is the stellar radius of the tachocline, and $\omega_c$ is
the angular velocity of the star beneath the tachocline. $G$ is a
measure of the strength of the negative buoyancy force in the
tachocline, and is sometimes referred to as the 'effective' gravity.
For adiabatic stratification this effective gravity is zero.
For each star we consider, we will use two values of $G$, one for the
radiative tachocline, and the other (much smaller value) for the
convective overshoot layer just above the radiative zone.

In solving equations (5-14) we use rigid top and bottom boundary
conditions, which mean that the boundaries are not allowed to
deform. There is no stress at the boundary, because our present
calculation is nonviscous. In future, when we attempt to do a
viscous calculation, we will have to implement a stress-free 
boundary condition explicitly, if we do not want to allow the
model to form boundary layers. For the magnetic field, we have
assumed perfectly conducting top and bottom boundaries in order
to confine magnetic field to the fluid shell. This is physically
reasonable for the bottom boundary, but somewhat artificial for
the top boundary, since we expect flux to escape upwards. To account
for this upward escape of flux is beyond the scope of the present
model. 

We use two independent numerical methods to analyze the
instabilities for $m=0$ as well as for $m>0$, namely a shooting method
and a spectral method. The details of the solution techniques have been
described in \citet{gdm2007} and in \citet{cally2003} for $m>0$
and in \citet{cdg2008} and \citet{dgcm2009} for $m=0$.

To find the values of $G$ and $\delta_s$ we use in our stability
calculations, we use the following parameters. Certain parameters
are chosen using the ZAMS stellar interior model of \citep{mjsm2007}.

\subsection{A typical F-star with 1.4 solar mass and 25 times solar 
rotator}

Most F-stars have a small convective core and a convective
outer envelope which is thinner than the Sun's. Our instabilities
will be applicable to the base of the outer convective envelope.
For a typical F star, we take a radius of $1.38 R_{\odot} = 9.59
\times 10^{10} {\rm cm}$ and a mass of $2.78 \times 10^{33} {\rm gm}$. 
We assume the outer $5\%$ of the star is the convective envelope; 
therefore the thickness of the  convection zone is $4.79 \times 10^4 
{\rm km}$. Assuming that the tachocline of a typical F-star is $7.5\%$ 
of the cz-thickness (like the Sun), its tachocline thickness is 
$3.6 \times 10^3 {\rm km}$. The density at the base of cz of this 
typical F star would be $8.85 \times 10^{-5}$ gm/cc. To estimate 
the pressure scale-height, we use a X=0.71, Y=0.27 H/He mixture, 
which leads to an atomic weight $\mu$ for the mixture of 0.71 * 0.5 + 
0.27 * 2= 0.90. We take the temperature at the base of the convection 
zone = $2.96 \times 10^5 {\rm K}$. Then for the universal gas constant, 
$R = 8.31 \times 10^7 {\rm erg}\, {\rm deg}^{-1}\, {\rm mole}^{-1}$ and 
gravity ($g$) at the base of the convection zone = $2.245 \times 10^4 
{\rm cm}^2\,{\rm s}^{-1}$, the pressure scale-height $(RT/\mu g) = 1.22 
\times 10^4$ km, the ratio of tachocline thickness 
to pressure scale-height) $\delta_s= 0.294$. From the table of values 
for our typical F star we take $|\nabla-\nabla_{ad}|= 10^{-1}$ 
and $10^{-4}$ for the radiative and overshoot tachoclines respectively, 
leading to $G=0.12$ and $1.2\times 10^{-4}$.

\subsection{A typical G star like the Sun}

Tachocline instabilities for solar type G-stars have been studied for a
wide range of $G$ values in \citet{gdm2007}. For comparison here we take
$G=10$ and $0.01$ for radiative and overshoot effective
gravities.

\subsection{A typical K-star with 0.75 solar mass and 1/2 times 
solar rotator}

Most K-stars have convective outer envelopes that are thicker
than the Sun's. For this case we take a radius of $0.6801 R_{\odot} = 4.7301
\times 10^{10} {\rm cm}$. The outermost $32.8\%$ is the convective envelope 
so the thickness of the CZ is $1.55 \times 10^5 {\rm km}$ and the tachocline 
thickness is $1.163 \times 10^4 {\rm km}$. The density at the base of the 
CZ is $1.61\, {\rm gm/cc}$. The mixture of $H$ and $He$ is the same as 
for the F star, for which $\mu = 0.8956$. In this star the temperature 
at the base of the convection zone = $2.8867 \times 10^6$ K and gravity 
($g$) at the base of the convection zone = $9.848 \times 10^4 {\rm cm}^2 
{\rm s}^{-1}$. Thus the pressure scale-height = $2.72 \times 10^4 {\rm km}$. 
Then $\delta_s = 0.428$. For this star $|\nabla-\nabla_{ad}|=5 \times 10^{-2}$ 
and $5 \times 10^{-9}$ for radiative and overshoot
tachoclines respectively. Therefore $G=2.5 \times 10^4$ for the radiative
tachocline and $2.5 \times 10^{-3}$ for the overshoot tachocline

\subsection{Magnetic field scaling}

Because of the different rotation rates and densities at tachocline depth
in F,G and K stars, the magnetic fields scale somewhat differently, but 
the effects of differing rotation and density tend to cancel each other 
out. In particular, in the Sun $a=1$ corresponds to $B=10^5 $ Gauss, 
while in F-stars $a=1$ is for $B=7.24 \times 10^4 $ Gauss and in 
K-stars $B=9.65 \times 10^4$ Gauss.

\section{Results}

Considering a typical case for the dynamo-generated toroidal band of 
$10^{\circ}$ latitudinal width,band-center located at $30^{\circ}$ 
latitude \citep{dg1999}, we compute the unstable eigen modes for 
this band in the parameter space of $-0.18 \le s \le 0.18$ and
$0.0 \le a \le 4$. We look for axisymmetric ($m=0$) as well as
non-axisymmetric ($m>0$) instabilities for various radial wavenumbers
$n$ in the overshoot and radiative tachocline of a star.

We sequentially present our results for a typical G-star, F-star
and K-star, for non-axisymmetric instabilities. Then
we give some examples of axisymmetric instabilities that could
occur. 

\subsection{Instabilities of solar and antisolar differential rotation
in G-stars}

Figure 1 displays the global MHD instability
for a G star such as the Sun, as a function of
the differential rotation parameter $s$ and
the peak toroidal field $a$. Frames (a) and (b)
show the growth rates for symmetric and antisymmetric
modes respectively, for an 'overshoot' tachocline
for which the effective gravity $G=0.01$.

We see that for large $a$, the growth rates are high
and nearly independent of the amount or even the
sign of the differential rotation present. Symmetric
and antisymmetric modes have virtually the same growth
rates. This implies that the nonlinear interaction between 
these modes is very likely to occur, triggering the flip-flop
mechanisms observed in many rapidly rotating stars \citep{fb2004,
fb2005}. We infer that for high $a$ the instability is governed
totally by the toroidal field.

\clearpage
\begin{figure}[hbt]
\epsscale{1.0}
\plotone{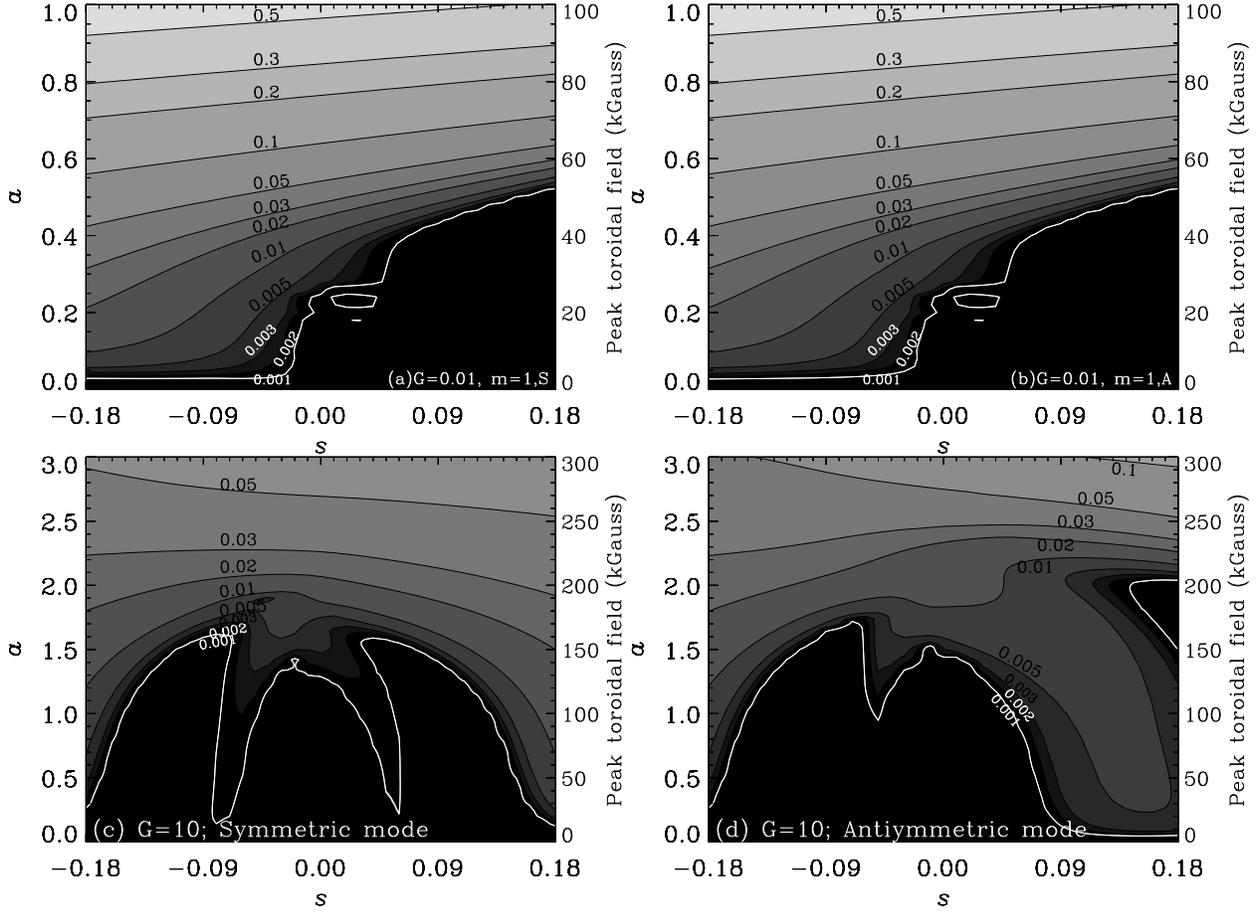}
\caption{Growth rate contours as function of $s$ and $a$
for a typical G-star. Frames (a) and (b) present growth rates for 
$m=1$ symmetric and antisymmetric modes with radial wavenumber, 
$n=1$ in the overshoot
tachocline with effective gravity, $G=0.01$, and frames (c) and (d) 
for the radiative tachocline ($G=10$). A growth rate of 0.01 
corresponds to an e-folding growth time of $\sim 1$ year in the
Sun. 
}
\label{Gstar-growth-rates}
\end{figure}
\clearpage

By contrast, for relatively weak toroidal fields, the amplitude
and sign of the differential rotation determines the growth
rate. Modes of both symmetries are unstable for much lower
toroidal fields for antisolar differential rotation than
for solar differential rotation. The difference in minimum
field strength needed for instability is about a factor of ten.
When applied to the tachocline of a G-star, this result should
mean that both antisolar differential rotation and tachocline
toroidal fields are kept at much smaller values by the instability
than they could be in a star with solar like differential rotation.
This may be one reason why so few cases of antisolar rotation
have been reported.

Frames (c) and (d) display growth rates for a radiative tachocline 
of a G-star. Here we again see that for high toroidal fields, the
instability is magnetically dominated and not strongly
dependent on differential rotation or mode symmetry. At low
toroidal field, antisolar differential rotation is no longer
unstable for much lower toroidal fields than is solar type
differential rotation. For both types of differential
rotation, larger toroidal fields are needed to excite instability
for lower differential rotation. For solar differential rotation,
the antisymmetric mode becomes unstable for lower toroidal field than
does the symmetric mode, especially for larger differential
rotation, while for antisymmetric differential rotation modes
of both symmetries are about equally unstable.

\clearpage
\begin{figure}[hbt]
\epsscale{1.0}
\plotone{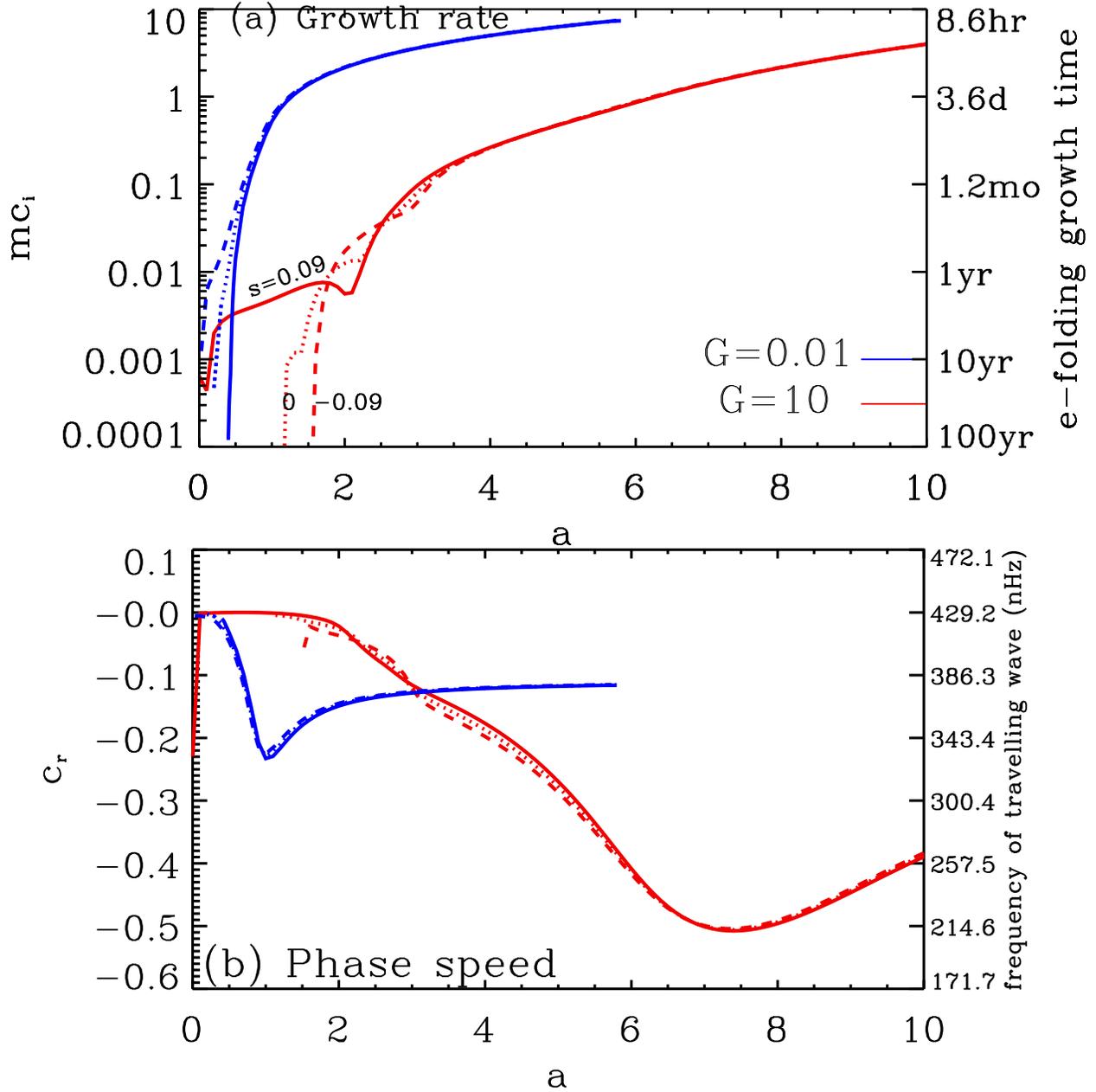}
\caption{Growth rate (frame a) and phase speed (frame b) curves as 
function of $a$ for selected differential rotation amplitudes, 
$S=-0.09$, 0 and 0.09, for a typical G-star. 
}
\label{Gstar-crci}
\end{figure}
\clearpage

The difference between low $G$ and high $G$ results comes from
the difference in disturbance structure. In particular,
for high $G$, the radial motion in disturbances is largely
suppressed by the negative buoyancy. It is nearly 2D 
(longitude-latitude) in pattern and is subject to relatively large 
longitudinal pressure gradients that produce longitudinal torques. 
By contrast, at low $G$ radial motions are much more prominent, 
and this allows the disturbances to more nearly conserve angular 
momentum locally. Such a disturbance is much easier to excite in 
a differential rotation profile that varies less with latitude. This 
favors antisolar over solar differential rotation. The larger the 
angular momentum gradient, the greater the perturbation centrifugal 
force pushing the displaced fluid elements back toward where they 
started. The perturbation $\mathbf{J \times B}$ force has to overcome 
this effect to create the instability, and less $\mathbf{J \times B}$ 
force is needed in the antisolar case.

Figure 2 shows a line plot of growth rate for the antisymmetric modes
depicted in Figure 1 for three differential rotations ($s=-0.09,0.,0.09$)
for $a$ from zero to ten. These results show the dominance of
the magnetic field in determining the growth rate as the field
is made stronger. At what field strength the toroidal field dominates 
depends on the effective gravity -- a stronger field is needed
to overcome the stronger negative buoyancy effects associated with
a larger effective gravity. We can see from Figure 2 that
the magnetically dominated instability is very powerful -- e-folding
growth times are measured in days or even hours compared to years
when the toroidal field is weak enough that differential rotation
dominates. 

Frame b) displays the phase velocity in longitude of the unstable 
modes, measured in fractions of the inertial frame of
core rotation rate, which is zero in the rotating
reference frame. We see that when the field is very weak,
unstable modes in the overshoot tachocline propagate
at approximately the rate of the core, which is
very close to that of the latitude of the band.
For extremely low toroidal field, modes in the
radiative tachocline propagate at a rate somewhat retrograde
relative to the core rate, typical of Rossby waves in
a rotating spherical shell. This is to be expected because
the perturbations in the radiative tachocline are nearly
two-dimensional (longitude-latitude). In both tachoclines the sign
and amplitude of the differential rotation has relatively
little effect on the propagation speed. For increased
toroidal field, the phase speed in both tachoclines is
still essentially independent of the differential rotation,
while differing from each other. This difference comes from the very 
different mode structure in the two tachoclines. Similar retrograde
propagation was found in previous analyses by \citet{gdm2007}.

\subsection{Disturbance planforms for radiative tachocline in G stars}

Examining the longitude-latitude structure of unstable
disturbances can help us understand the physics of the
instability seen in Figures 1 and 2. Figures 3, 4 and 5
display for G star cases the longitude-latitude patterns
of disturbance velocities, magnetic fields and fluid pressure.
Figure 3 is for antisymmetric disturbances in a radiative
tachocline for relatively high toroidal field parameter
$a$ of 1.0 and 2.5. The left column of panels (a,c,e)is the
antisolar case, the right column (panels b,d,f) the solar case.

We see in panels a) and b) that the perturbation magnetic
fields, represented by arrows, are tightly confined to the 
neighborhood of the toroidal band, as we should expect. These 
field lines are essentially closed in an oval, meaning that 
there is very little perturbation field that is vertical. We see 
further that the perturbation field vectors are oriented clockwise 
around the highs (red contours) in perturbation fluid pressure and 
counterclockwise around the lows (blue contours). This is the 
opposite orientation to the so-called geostrophic balance between 
velocity vectors and fluid pressure. In that case there is a near 
balance of the Coriolis and pressure gradient forces. Here there is 
a tendency for the perturbation $\mathbf{J \times B}$ and fluid 
pressure gradient forces to balance. For high field strengths
in the neighborhood of the toroidal band, Coriolis forces
are less important.

The direction of the arrows implies that the total
horizontal field (undisturbed, positive, toroidal band plus
perturbations) is displaced in latitude toward where the
perturbation east-west field is positive (toward the right)
and away from where it is negative (pointed toward the left).
Given the position of high (red) and low fluid perturbation pressure
(blue) centered at the latitude of the undisturbed band, we infer
that the peaks and troughs of total perturbation pressure (fluid plus
magnetic) are displaced in latitude in the same way as the total field.

The perturbation magnetic patterns are very similar for antisolar and
solar differential rotation. But away from the band, the fluid
pressure patterns are rather different. This is where hydrodynamic
processes are dominant. Panels c) and d) display the perturbation
velocity vectors for antisolar and solar cases. We see that away from
the latitude of the toroidal band, the flow is nearly geostrophic--
clockwise around the highs of fluid pressure, counterclockwise
around the lows. But within the band, the flow easily crosses the
fluid pressure contours. Here it is being driven across by the
perturbation $\mathbf{J \times B}$ force, carrying field with it to 
generate the poleward and equatorward displacements of total field.

\clearpage
\begin{figure}[hbt]
\epsscale{1.0}
\plotone{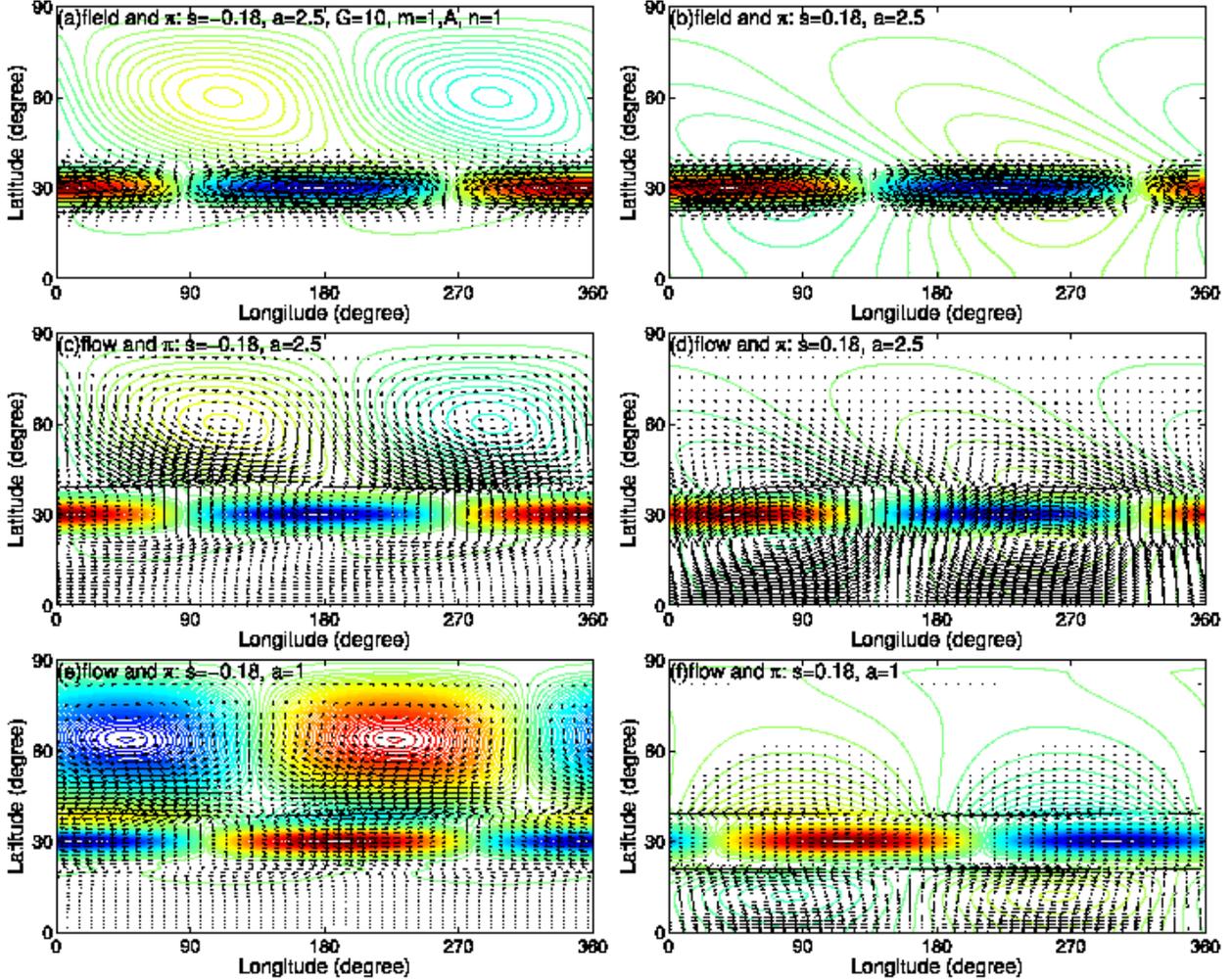}
\caption{Left and right columns respectively show eigen functions
for antisolar ($s=-0.18$) and solar-like ($s=0.18$) disturbance 
patterns in latitude-longitude plane, for a typical radiative
tachocline. Magnetic vectors superimposed on color contours of 
total pressure perturbations are presented in top two frames, (a) 
and (b), for $a=2.5$. Horizontal velocities (black arrows) 
superimposed on fluid pressure perturbations, $\pi$, (color contours) 
have been plotted for $a=2.5$ (frames (c) and (d)) and for $a=1$ 
(frames (e) and (f)).
}
\label{planform_radiative}
\end{figure}
\clearpage

The 'tilts' of the velocity vectors from the E-W direction, seen
most strongly on the poleward side of the band, are of particular
interest. In both antisolar and solar cases they imply angular
momentum transport toward the equator. In the antisolar case, this
implies angular momentum transport down the angular velocity
gradient, which feeds kinetic energy to the disturbances
away from the band, and would reduce the differential rotation.
By contrast the tilts in the solar case imply angular momentum
transport up the gradient, which would actually increase the
differential rotation. The energy for this increase comes from
the toroidal field. So there will be further tendency for
antisolar differential rotation to be destroyed by this instability.
In the solar case, these high latitude hydrodynamic perturbations
must be driven by the low latitude $\mathbf{J \times B}$ forces and 
associated fluid pressure gradients, because they are giving up energy
to the differential rotation.

Panels e) and f) display the velocity and pressure perturbations
for a weaker peak toroidal field parameter $a=1$, which is much
closer to the value for onset of this instability. Here we see the tilts
have largely disappeared. The velocity patterns still show
the near geostrophic balance, but are largely present only in high
latitudes in the antisolar case, mostly in equatorial latitudes
in the solar case.

\subsection{Disturbance planforms for overshoot tachocline in G star}

Figures 4 and 5 show planforms for unstable disturbances
in the overshoot tachocline of a G-star, with effective gravity
$G=0.01$, for modes that are symmetric about the equator.
Figure 4 is for a strong toroidal field case, namely $a=0.6$, for
both antisolar and solar type differential rotations. Figure 5
is for a weak field case, $a=0.1$, for which, from Figure 1,
only the antisolar case is unstable.

\begin{figure}[hbt]
\epsscale{1.0}
\plotone{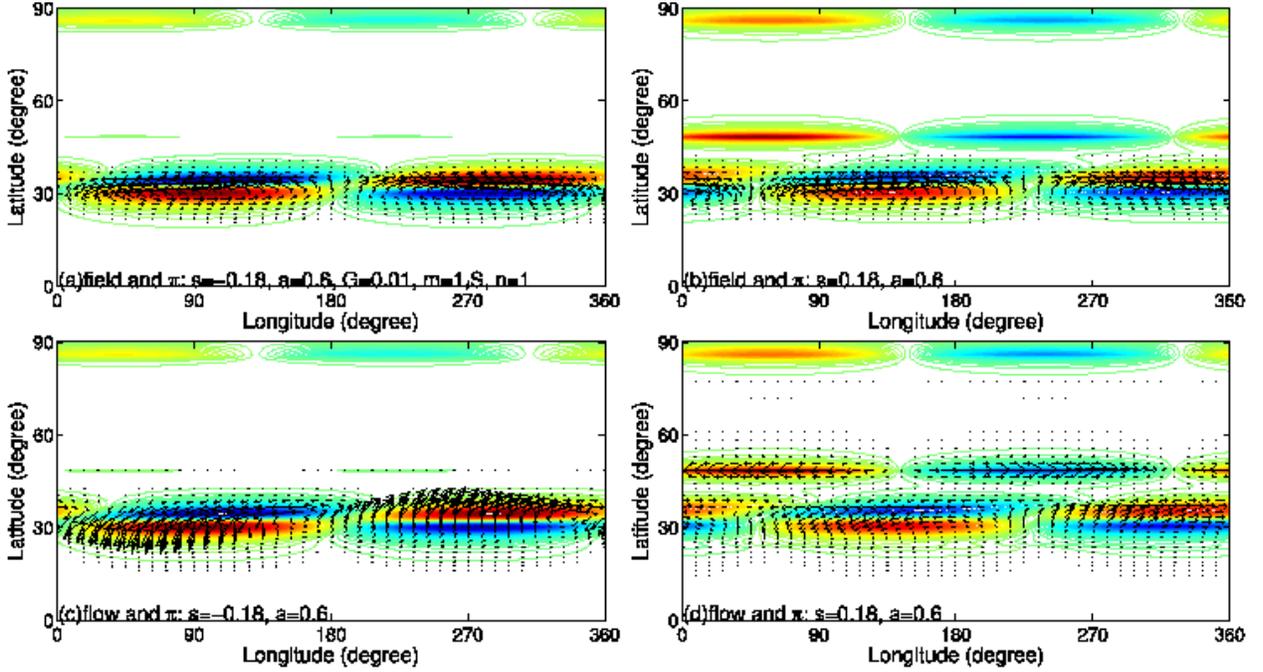}
\caption{Left and right columns respectively show eigen functions
for antisolar ($s=-0.18$) and solar-like ($s=0.18$) disturbance 
patterns in latitude-longitude plane, for a typical overshoot
tachocline. Magnetic vectors superimposed on color contours of 
total pressure perturbations are presented in top two frames, (a) 
and (b), for $a=0.6$. Horizontal velocities (black arrows) 
superimposed on fluid pressure perturbations (color contours) 
are presented in bottom two frames (c and d).
}
\label{planform_ov_hia}
\end{figure}

For $a=0.6$ in Figure 4, we see the perturbation fluid pressure
and velocity structures (frames c) and d))are very different from 
the case of the radiative tachocline. Peaks and troughs of fluid 
pressure occur at nearly the same longitude, and there are two 
latitudinally narrow additional fluid pressure perturbations on the 
poleward side of the band, one near it and the other near the 
poles. Most of the magnetic perturbations (frames a) and b)) of 
both signs occur on the poleward side of the toroidal band. High 
fluid pressure is still on the left of the perturbation field arrows, 
low pressure on the right, but the perturbation fields are not 
closed ovals. Rather there are strong points of horizontal 
convergence and divergence of field. This means the vertical 
fields are quite significant. Similarly, the flow fields are 
not simply counterclockwise around lows and clockwise around 
highs. Instead there is much flow both up and down the fluid
pressure gradients, and substantial areas of horizontal convergence
and divergence. This implies the vertical motions are substantial.

Overall, the flow pattern is one in which, in each half wavelength 
in longitude, there is a local 'roll' with E-W axis, the sense of which
reverses in the next half wavelength. Within each roll, flow
toward the equator acquires a negative component relative to the 
rotating frame, flow toward the pole, a positive component. This 
means that locally the circulation in the roll is tending to conserve 
angular momentum, only weakly opposed by any longitudinal fluid pressure 
torques. In the radiative tachocline with high effective gravity, 
the strong negative buoyancy forces strong longitudinal fluid pressure
torques that are balanced by the Coriolis force on the nearly 
horizontal motion. The limited latitudinal extent of the velocity 
and fluid pressure perturbations in the overshoot case is preferred 
because that minimizes the stabilizing effect of the perturbations 
conserving angular momentum.

\clearpage
\begin{figure}[hbt]
\epsscale{0.7}
\plotone{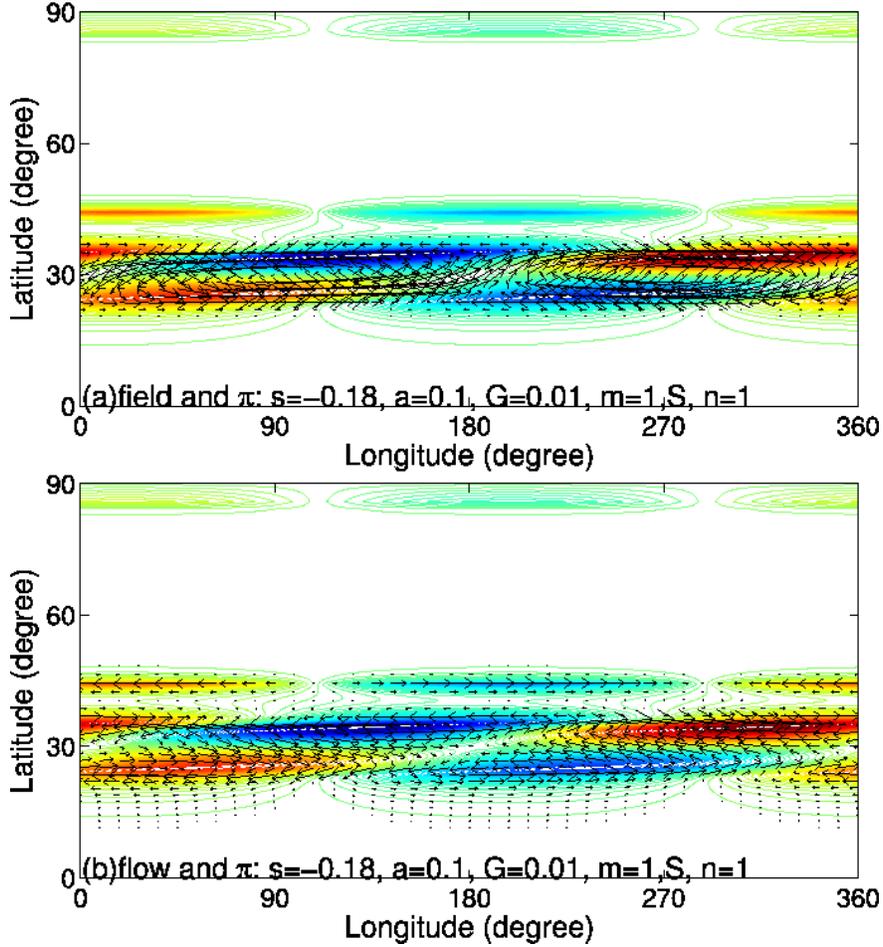}
\caption{Left and right columns respectively show magnetic and flow 
vectors for antisolar ($s=-0.18$) disturbance patterns in 
latitude-longitude plane, for a typical overshoot tachocline and 
for a low field strength, $a=0.1$. Color contours for fluid 
pressure perturbations have been superimposed on arrow-vectors in 
both frames. Solar-like differential rotation is stable for such
a low field strength (see Figure 1(b)).
}
\label{antiplanform_ov_loa}
\end{figure}

Figure 5 for the weak field antisolar case shows an even more
complex perturbation pattern. Here the flow again (frame b)) has 
a tendency to be geostrophic, with counterclockwise flow around the low
fluid pressure and clockwise around the high. This happens because the
magnetic field is too weak to force the flow to do otherwise. Now
the perturbation field peak vectors tend to coincide with the peaks
in fluid pressure. At this low field, hydrodynamics is more dominant
everywhere. The energy for instability in this case is coming
primarily from the differential rotation. The tilt in the velocity
vectors is such that flow toward the pole has a strong component
(in the rotating frame) opposite to the direction of rotation,
flow toward the equator, a component in the direction of rotation.
These tilts imply a Reynolds stress that transports angular momentum
from high latitudes with high angular velocity to low latitudes
with low angular velocity, thereby extracting kinetic energy from
the antisolar differential rotation. The same tilt in the
case of solar type differential rotation would do the opposite,
and perturbations in this case would not grow.

\subsection{Instabilities of solar and antisolar differential rotation
in F-stars}

Figure 6 displays growth rates for overshoot (frames a and b) and
radiative (frames c and d) tachoclines in an $F$ star whose convection
zone is about 5\% of its radius. The dimensionless vertical scale
for the peak toroidal field $a$ is the same as for G stars shown 
in Figure 1, but the dimensional fields on the right hand scale are
smaller for the same $a$. In general, the effective gravity of the
F-star's overshoot tachocline is lighter than that of G and K stars.
Although the nonhydrostatic-Boussinesq and the 
hydrostatic-nonBoussinesq models produce very close results except
for $G \approx 0$, it often becomes numerically difficult to compute 
the growth rates in a hydrostatic model with a very light effective
gravity. The results presented in frames (a) and (b) of Figure 6
are obtained using Boussinesq system.

Comparing frames of Figure 6 to the corresponding ones in Figure 1, we
see that the instability for F stars is qualitatively similar to
those for G stars, but there are substantial quantitative differences.
For high toroidal fields the instability is essentially independent
of differential rotation, while for low toroidal fields in the overshoot
tachocline antisolar differential rotations are unstable for much lower 
toroidal fields than are solar-type differential rotations. In the 
radiative tachocline of an F star, anti-solar and solar type differential 
rotations are about equally unstable, similar to G stars.

In both types of tachocline, instability sets in at substantially lower 
toroidal field (both dimensional and dimensionless) than in G stars, and 
for the same differential rotation and dimensionless toroidal field, the 
growth rates are much larger than those for G stars. These differences 
arise from the fact that the effective gravities in F star overshoot
and radiative tachoclines are substantially smaller than those of their
G star counterparts.

\clearpage
\begin{figure}[hbt]
\epsscale{1.0}
\plotone{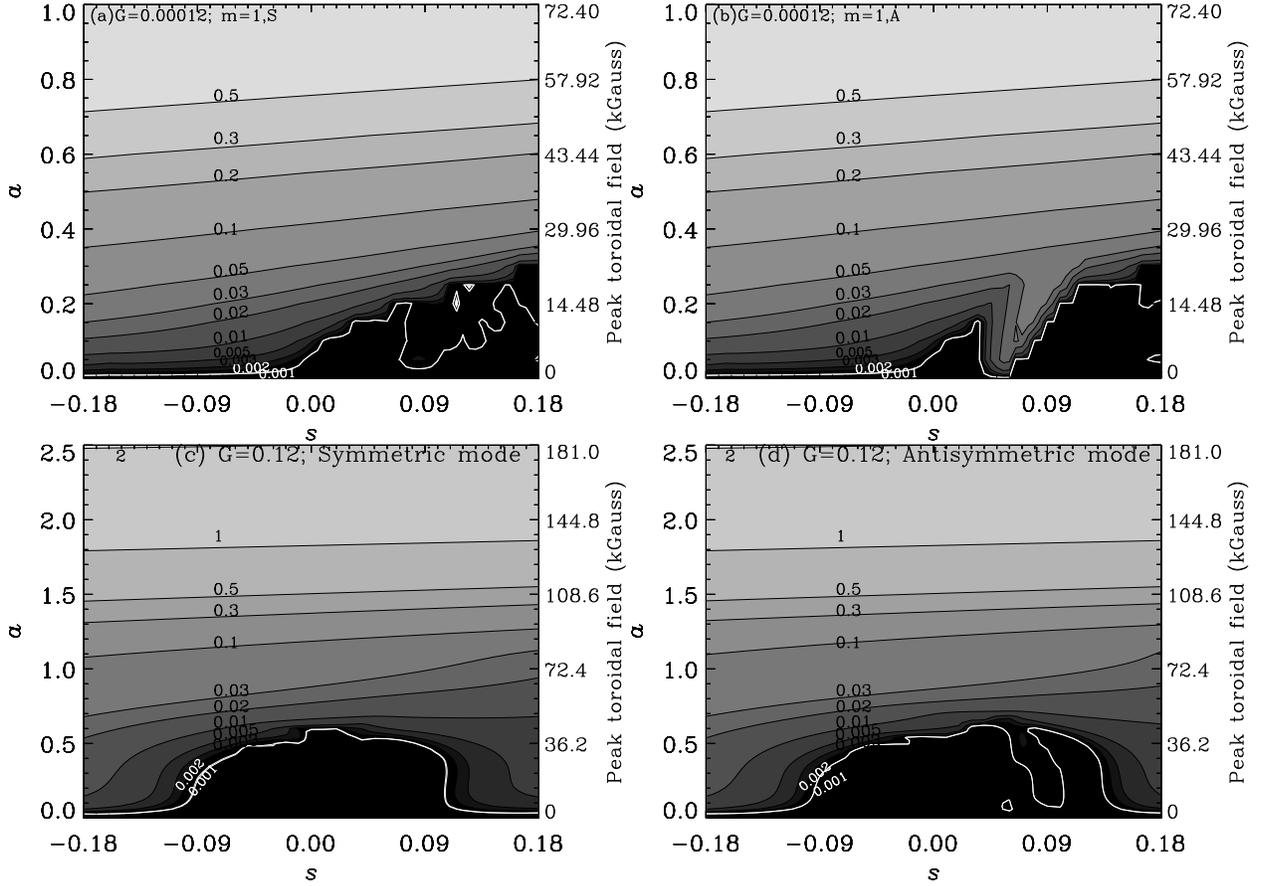}
\caption{Growth rate contours as function of $s$ and $a$
for a typical F-star. Frames (a) and (b) present growth rates for 
$m=1$ symmetric and antisymmetric modes with radial wavenumber, 
$n=1$ in the overshoot
tachocline with effective gravity, $G=0.00012$, and frames (c) and (d) 
for the radiative tachocline ($G=0.12$). 
A growth rate of 0.01 here corresponds to an e-folding growth time 
of $\sim 15$ days in a typical F star rotating 25 times faster
than the Sun. 
}
\label{Fstar-growth-rates}
\end{figure}
\clearpage

These results imply that in F stars even relatively weak toroidal 
fields will cause instability that should largely wipe out the 
differential rotation (nonlinear calculations would be needed to test 
this conclusion). This inference implies that substantial differential 
rotation would be very hard to maintain in the tachocline of an F star, 
particularly if it is of antisolar type. Since F star convection zones 
are thin and their tachoclines are not far below their photospheres, 
we infer this instability could severely limit the amplitude of surface 
differential rotation in F stars. 

\subsection{Instabilities of solar and antisolar differential rotation
in K-stars}

Figure 7 displays growth rates for overshoot (frames a),b))
tachoclines of K-stars, again for
disturbances of both symmetries about the equator.
We see that in the overshoot case, the result is
qualitatively similar to that for G stars. For
weak field only antisolar differential rotations lead to
instability for low peak field in the toroidal band.
Instability occurs for somewhat lower toroidal field
for all differential rotations, because the effective
gravity $G$ in K-star overshoot tachoclines is about
25\% that for G stars. Therefore in overshoot tachoclines
for all three classes of stars, antisolar differential rotations
with banded toroidal fields are unstable for much smaller
peak toroidal fields than are solar type differential rotations.

\begin{figure}[hbt]
\epsscale{1.0}
\plotone{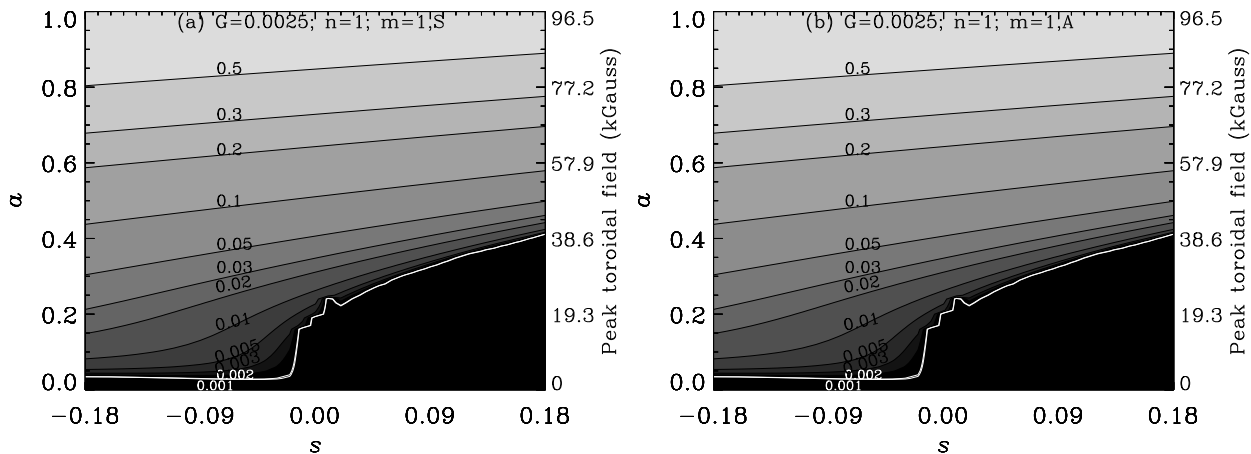}
\caption{Frames (a) and (b) present growth rate contours in the
overshoot tachocline of K star for $m=1$ symmetric and antisymmetric
modes, with radial wavenumber, $n=1$, and effective gravity, $G=0.0025$. 
A growth rate of 0.01 here corresponds to an e-folding growth time 
of $\sim 2$ years in a typical K star rotating 0.5 times slower
than the Sun. 
}
\label{Kstar-growth-rates}
\end{figure}

\clearpage

Instability in radiative tachoclines of K stars is mostly
present only for $s>0.10$ or so. These instabilities proved to be 
much more difficult to calculate than for F and G stars, using
both shooting and spectral methods. The result obtained shows a 
lot of fine structure in growth rates that are almost all very 
weak, especially for antisolar rotation -- we do not display them here. 
However, we note that for solar type rotation with $s>0.10$ more accurate
solutions can be obtained, and the results correspond to growth rates
of typically up to 0.006 in the antisymmetric case, and about double that
for symmetric instabilities. Growth rates are apparently much smaller 
than this for antisolar rotations. There seems to be little dependence
on $a$ in all cases, suggesting the instabilities are predominantly
hydrodynamic.
The noisy eigen values in the parameter space could be due to the 
extremely high effective gravity of K star, which makes the equations 
solved very stiff by taking them to the computational limitation 
of the methods used. 

\subsection{Axisymmetric instabilities of solar and antisolar 
differential rotation in F, G and K-stars}

Instability of axisymmetric ($m=0$) modes is
almost impossible to excite in radiative tachoclines
of F,G or K stars, because vertical displacements are required
and the negative bouyancy force from the strongly
subadiabatic temperature gradient opposes such displacements.
In the overshoot tachoclines of such stars, however, it
is possible to excite axisymmetric modes. Figure 8 gives the result
for these cases, for modes that are symmetric about the equator.
Antisymmetric modes give very similar results.

We see that instability occurs essentially independent of the 
differential rotation for $a$ values significantly above 0.7 
(F-stars), 0.9 (G-stars) and 0.8 (K-stars). This is a purely 
magnetic instability in which the differential rotation plays 
very little role. So does the symmetry of the modes. In general, 
the instability for non-zero $m$ in overshoot tachoclines described 
earlier in this paper occurs for much lower values of toroidal 
field parameter $a$. So we infer that these $m=0$ modes of 
instability are less important for these classes of stars, though
there is a weak tendency for instability to set in at lower field
strengths for antisolar compared to solar differential rotation. 

\clearpage
\begin{figure}[hbt]
\epsscale{0.55}
\plotone{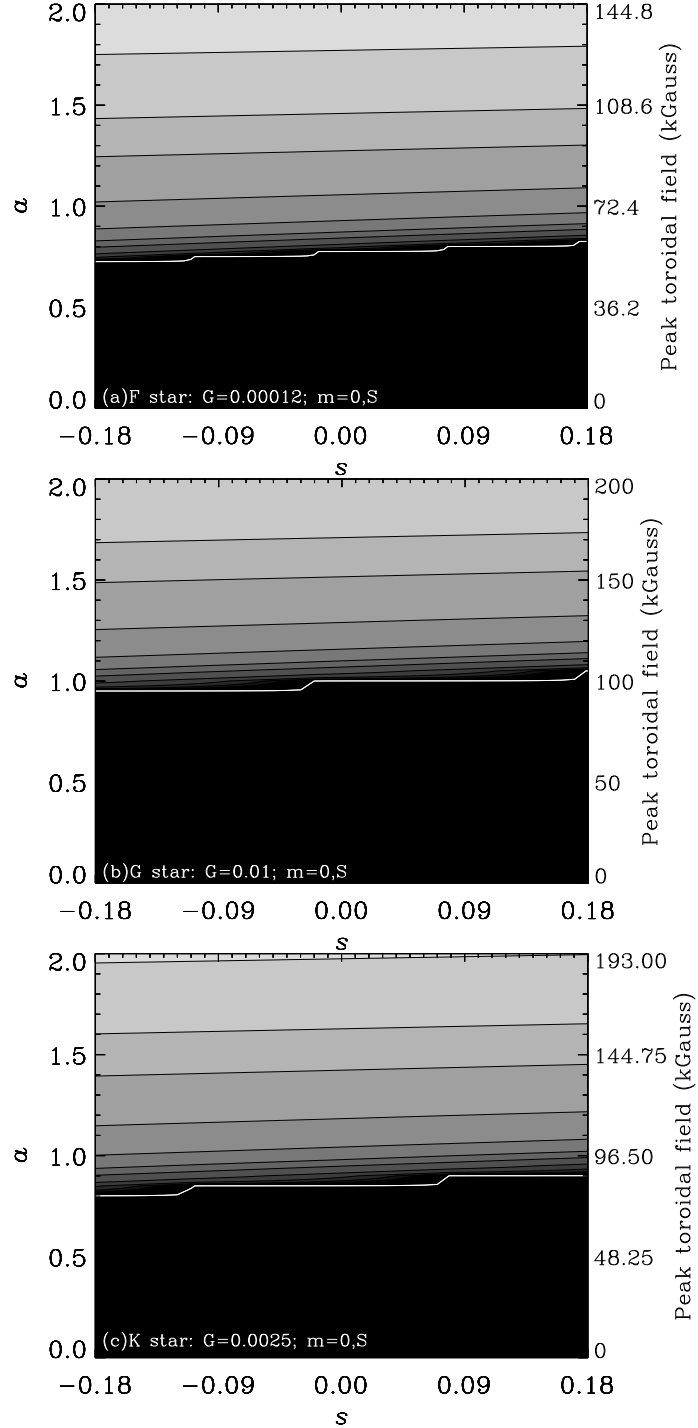}
\caption{Frames (a), (b) and (c) present growth rate contours for $m=0$
instabilities respectively in the overshoot tachoclines of F, G and K 
stars for $m=1$ symmetric and antisymmetric modes, with radial wavenumber, 
$n=1$. White contour represents the stability boundary designated by
the growth rate value 0.001. Gray-scale filled contours are in 
roughly logarithmic intervals (0.01, 0.02, 0.03, 0.05, 0.1 ...).
}
\label{FGK_m0}
\end{figure}
\clearpage

\subsection{Instabilities of solar and antisolar differential 
rotation with higher radial wavenumbers}

Although we focussed primarily on instability with radial
wave numbers 1 corresponding to half a wavelength across the
layer, our instability equations allow modes of
any radial wavenumber $n$, and we have explored some cases
with $n=10$ and $n=100$ (plots not presented).
In general the
higher the $n$ chosen, the shorter are the radial
displacements. Shorter displacements imply reduced
resistance by any negative buoyancy force. For overshoot
tachoclines the low effective gravity implies there is little
buoyant resistance to vertical motions. So the growth rate
of unstable modes is nearly independent of $n$. 

For radiative tachoclines, the strong negative buoyancy due to
the high effective gravity $G$ suppresses the radial motions. So 
it becomes easier to excite unstable modes with higher $n$,
because modes with higher $n$ contain smaller radial displacements,
and hence should feel less resistance from the strong negative 
buoyancy. Therefore
they could be more unstable, or be unstable for parameter
values for which $n=1$ modes are stable. 

Our results for radiative tachoclines in F, G, and K stars
represent the minimum amount of instability in the system which
is for $n=1$ -- there could be even more for $n>1$. If diffusion 
were included in the model, which is certainly present in stellar 
tachoclines, modes with higher $n$ would become less unstable or 
even stable. Higher $n$ modes will also be less unstable if 
the toroidal bands are twisted \citep{fllfp1999}. 

%\begin{figure}[hbt]
%\epsscale{1.0}
%\plotone{fig_n1fgkm0.eps}
%\caption{Frames (a), (b) and (c) present growth rate contours for $m=0$
%instabilities respectively in the overshoot tachoclines of F, G and K 
%stars for $m=1$ symmetric and antisymmetric modes, with radial wavenumber, 
%$n=1$. 
%}
%\label{FGK_m0}
%\end{figure}

\section{Concluding comments}

From analyses of global MHD instabilities in 3D thin-shell models
of the solar/stellar tachoclines, we find that solar-like and antisolar 
type latitudinal differential rotations with coexisting toroidal bands
are unstable. Antisolar differential rotation is, in general, more 
unstable than a solar-like one in all F, G and K stars -- the pole
rotating one or two percent faster than the equator can make the
latitudinal differential rotation unstable in weakly magnetized
overshoot tachoclines, although it requires much stronger magnetic
fields to be unstable in the radiative tachoclines in G and K stars.

High effective gravity in the radiative tachoclines largely suppreses
the radial motions by the negative magnetic buoyancy and therefore,
it is difficult to grow the perturbations. By contrast, low effective
gravity allows the radial motions to grow, and more so when the variation
in differential rotation in latitude is smaller. Thus antisolar type
differential rotation becomes more unstable than solar type ones, because
less $J\times B$ force is needed to overcome the perturbation
centrifugal force that tries to stabilize the displacement.

The eigen functions for the disturbances with solar-like and antisolar
type differential rotations have similarities in the high $G$ case, but 
have important differences in the low $G$ case. In the high $G$ case, 
the velocity and pressure perturbations are in nearly geostrophic 
balance and horizontal divergences in flow and field are very small. 
The E-W tilts of the velocity vectors indicate angular momentum transport
towards the equator, but this means that the instability tries to
destroy the antisolar type gradient in latitudinal differential 
rotation, whereas it builds the solar type gradient. However, in
both cases, the primary energy source for the disturbances 
is the toroidal field, so for high $G$, solar and antisolar 
differential rotations require a similar amplitude of toroidal
field to become unstable.

For low $G$, by contrast, the eigen functions have more radial
motions and fields and associated horizontal divergences. For weak
toroidal fields, the hydrodynamics tend to dominate at all latitudes
including at the locations of the toroidal bands. The energy for
the growth of the disturbances is primarily coming from the differential
rotation. The E-W tilts in horizontal velocity vectors for 
antisolar type differential rotations imply angular momentum
transport towards the equator, which takes kinetic energy out of
the differential rotation. Disturbances with the same tilts would
actually reinforce a solar type differential rotation, damping the
disturbances. This is why instability for solar type differential
rotations does not occur until a substantially higher toroidal
field is included.    

For F stars, antisolar latitudinal differential rotation of only 
a few percent pole-to-equator amplitude is unstable in both the 
overshoot and radiative tachoclines, because the effective gravity 
is much smaller in F stars compared to G and K stars. Therefore, 
antisolar differential rotation is unlikely to be large in F stars.
By contrast, in K stars, antisolar differential rotation is stable 
in the radiative tachocline but unstable in the overshoot tachocline. 
This implies that we should find more K stars with antisolar 
differential rotation compared to F stars, but the observations 
have so far indicated the opposite, namely no K stars and about 
half a dozen of F stars are found to have antisolar differential 
rotations. 

We might be able to detect more antisolar K stars in future. 
One important point we should restate again is that the stability 
analyses we are performing is in the tachoclines of these stars, 
and stellar differential 
rotations are generally observed at the surface. The deeper convection
zone in the K stars might mean that the surface is not reflecting
the tachocline differential rotation pattern, whereas a very thin
convection zone in F stars may be relfecting a 1 or 2\% stable
antisolar tachocline differential rotation at the surface. In any
case, antisolar stars are found to have a very small amount of
positive pole-to-equator differential rotation.

In this paper, we did not attempt to understand what builds a
solar-like or an antisolar type differential rotation, but once
they are built, we investigated their stability to disturbances
with various longitudinal and radial wave numbers. We focussed on
a $10^{\circ}$ toroidal band placed at $30^{\circ}$ latitude in
the tachoclines of these stars, but other latitude locations of
the bands and band-widths can be studied for antisolar differential
rotations. Full 3D models have been explored by \citet{asr2007} and 
\citet{zls2003} for the stability analysis of the solar-like 
differential rotation, and antisolar case is yet to be explored.

\acknowledgements

We thank Peter Gilman for reviewing the manuscript and for many
helpful discussions. We also thank Keith MacGregor for
giving us the stellar structure data computed from his ZAMS stellar
interior model and Travis Metcalfe for helpful discussions at the
early stage of the model-calculations. We extend our thanks to an anonymous 
referee for a thorough review of the previous version of the manuscript 
and for his/her many helpful comments, which have helped improve this 
paper significantly. This work is partially supported by NASA's Living 
With a Star program through the grant NNX08AQ34G. The National Center 
for the Atmospheric Research is sponsored by the National Science 
Foundation.

%%%%%%%%%%%%%%%%%%%%%%%%%%%%%%%%%%%%%%%%%%%%%%%%%%%%%

\end{document}